\documentclass[conference]{IEEEtran}
\IEEEoverridecommandlockouts
\usepackage{cite}
\usepackage{amsmath,amssymb,amsfonts}
\usepackage{algorithm}
\usepackage{algorithmicx}
\usepackage{algpseudocode}
\usepackage{graphicx}
\usepackage{textcomp}
\usepackage{xcolor}
\usepackage{longtable}
\usepackage{hyperref}
\usepackage{multirow}
\usepackage{pifont}
\usepackage{booktabs}
\usepackage{enumitem}
\usepackage{epstopdf}
\usepackage{epstopdf}

\makeatletter

\def\ps@IEEEtitlepagestyle{%
  \def\@oddfoot{\mycopyrightnotice}%
  \def\@evenfoot{}%
}
\def\mycopyrightnotice{%
}

\def\BibTeX{{\rm B\kern-.05em{\sc i\kern-.025em b}\kern-.08em
    T\kern-.1667em\lower.7ex\hbox{E}\kern-.125emX}}

    \newcommand{\blue}[1]{\textcolor{black}{#1}}
        
    \definecolor{wjcolor}{rgb}{0.6, 0.4, 0.8}

\begin{document}

\title{\blue{Electric Vehicle Charger Infrastructure Planning: Demand Estimation, Coverage Optimization Over an Integrated Power Grid}
% \thanks{}
}

% \author{
%     \IEEEauthorblockN{Jingbo Wang}\IEEEauthorrefmark{1},
%     \IEEEauthorblockN{Harshal Kaushik}\IEEEauthorrefmark{1},
%     \IEEEauthorblockN{Roshni Anna Jacob}\IEEEauthorrefmark{1},
%     \IEEEauthorblockN{Jie Zhang}\IEEEauthorrefmark{1},
%     \IEEEauthorblockA{
% \textit{The University of Texas at Dallas}\\
% Richardson, Texas, U.S. \\
% jingbo.wang@utdallas.edu}}

\author{\IEEEauthorblockN{Harshal D. Kaushik, \textit{Member, IEEE,} Jingbo Wang, \textit{Graduate Student Member, IEEE,} \\ Roshni Anna Jacob, \textit{Member, IEEE,} Jie Zhang, \textit{Senior Member, IEEE}}
    \IEEEauthorblockA{The University of Texas at Dallas,  Richardson, Texas 75080, USA \\
    \ Email: jiezhang@utdallas.edu}
}

\maketitle

\begin{abstract}
For electrifying the transportation sector, deploying a strategically planned and efficient charging infrastructure is essential. This paper presents a \blue{two-phase} approach for electric vehicle (EV) charger deployment that integrates spatial point-of-interest analysis and maximum coverage optimization \blue{over an integrated spatial power grid}. Spatial-focused studies in the literature often overlook electrical grid constraints, while grid-focused work frequently considers statistically modeled EV charging demand. \blue{To address these gaps, a new framework is proposed that combines spatial network planning with electrical grid considerations. This study approaches EV charger planning from a perspective of the distribution grid, starting with an estimation of EV charging demand and the identification of optimal candidate locations.} It ensures that the capacity limits of newly established chargers are maintained within the limits of the power grid. This framework is applied in a test case for the Dallas area, integrating the existing EV charger network with an 8500-bus distribution system for comprehensive planning.
\end{abstract}


\section{Introduction}
The widespread adoption of electric vehicles (EVs) depends significantly on the expansion of the charging infrastructure network. %Increasing the number of charging points can substantially alleviate the range anxiety associated with EVs' relatively limited driving distances. 
Establishing charging station infrastructure practically involves three key stages: first, estimating the EV charging demand within the target area; second, identifying optimal locations for new chargers; and third -- often overlooked -- ensuring that the charger capacities do not exceed limits that could cause power grid violations. %This ensures that all demand within the area is met while also maintaining the manageable loads on the power grid. 

% Steps in designing the infrastructure for EV charging:
% \begin{itemize}
%     \item Estimate the EV charging demands.
%     \item Maximum coverage problem for optimum location.
%     \item Estimation of chargers capacity using DSS simulations.
% \end{itemize}
% Next, these steps are discussed. Existing literature, gaps, and our contributions.  

A common approach in the literature for estimating the EV charging demand involves modeling EV battery usage and simulating energy consumption under various traffic scenarios. However, obtaining realistic data on actual EV charging demands in specific areas is often challenging  \cite{cai2014siting}. Some studies predict charging demands based on vehicle ownership, travel behavior, and household travel surveys \cite{sadeghi2014optimal,shahraki2015optimal,chen2013locating,xi2013simulation,hu2018analyzing, kontou2019understanding, vazifeh2019optimizing}.
% For instance, one study explored the relationship between daily activities and charging demands for private vehicles \cite{kontou2019understanding}.  %\href{https://www.sciencedirect.com/science/article/pii/S0968090X18305539?via%3Dihub}{8} 
% Another work used cellular data to construct trajectories in Boston \cite{vazifeh2019optimizing}. %\href{https://www.sciencedirect.com/science/article/pii/S0965856417300010?via%3Dihub}{9}. 
The main issue with these simulation-based approaches is their computational expense, which makes them difficult to scale. 

Given these scalability challenges, data-driven approaches have gained popularity in recent years. These methods typically divide the area into discrete spaces, extract driver's patterns from the travel mobility data, and estimate public charging demands for each cell \cite{vazifeh2019optimizing,kontou2019understanding,dong2019electric}.
Some studies consider the proximity of demand from neighboring regions \cite{wagner2013optimal,tu2016optimizing,dong2019electric}, while others argue that the charging demand depends on the specific characteristics of each region \cite{yi2022electric}. In these approaches, each discrete region is considered as a node, and nodes are ranked based on their unique features. %We will use a similar methodology to rank the discrete regions and estimate the demands for EV charging. 
    % \subsection{Facility location problem for selecting the location}
% In the literature, the flow capturing location model is a major tool used for optimizing transportation networks \cite{hodgson1990flow, jingbo_hdk2024, hdk_raj2024}, 
% where EV flows occur along origin-destination pairs. The goal is to maximize the volume of charged vehicle flows \cite{kadri2020multi,erdougan2022establishing}. 
% However, this approach faces scalability issues again for larger networks\cite{hdk_ACC, hdk_TAC}. Approximate methods must handle complex factors such as traffic flow integration, EV penetration levels, charging demand uncertainty, and time-related considerations such as charging duration, waiting, and driving distance, adding complexity to the problem formulation. Placing EV charging stations introduces significant power demands on the grid, which can fluctuate due to unpredictable charging needs, affecting the distribution of active and reactive power. To ensure reliability and safety, voltage and current levels at any bus must remain within certain ranges to prevent network overloads \cite{Jingbo_raj2023}. Alternative grid reinforcement approaches discussed in the literature to address these challenges \cite{chen2023research,tao2022adaptive}. 
\blue{Flow-capturing location model is a major tool in the literature %where the goal is to maximize the volume of charged vehicle flows. %\href{https://www.sciencedirect.com/science/article/pii/S0305054820300058}{2}, \href{https://www.sciencedirect.com/science/article/pii/S0038012121001191}{3}. 
% However, this approach 
that faces scalability issues  for larger networks\cite{hdk_ACC, hdk_TAC, kadri2020multi,erdougan2022establishing, hodgson1990flow, jingbo_hdk2024, hdk_raj2024}.}

From the power grid perspective, placing the EV charging stations introduces significant power demands on the grid, which can fluctuate due to unpredictable charging needs, affecting the distribution of active and reactive power. \blue{To ensure reliability and safety, voltage and current levels at any bus must remain within certain ranges to prevent network overloads \cite{Jingbo_raj2023, chen2023research,tao2022adaptive}.}% Alternative grid reinforcement approaches discussed in the literature to address these challenges \cite{chen2023research,tao2022adaptive}.

\blue{There are several gaps in the current literature. The primary challenge is that spatial-focused approaches often overlook power grid considerations and power grid-centric methods overlook spatial aspects. Secondly, estimating EV charging demand is not only complex but also difficult to scale up. %, as previously discussed.
To address this, recent works such as \cite{dong2019electric, yi2022electric, wagner2013optimal} have explored data-driven approaches for EV charging demand prediction using urban informatics and mobility data. Building on this, a multidisciplinary two-phase approach is developed in this work. 
%for prediction and coverage optimization.
%A data-driven method is employed to predict EV charging demands, and a maximum coverage optimization model is solved to identify optimal charger locations. OpenDSS simulations and power flow calculations are performed to avoid potential overloading and ensure sustainable charger deployment in alignment with the distribution network. 
The contributions of this work are as follows:
\begin{itemize}[left=-0.3pt]
    % \item We use a data-driven point-of-interest methodology to predict the EV charging demand. We divide the target area into a grid and estimate the EV charging demand for representative cells. Using the Google Maps dataset, we scrape data to predict charging demand.
    \item A data-driven point-of-interest methodology is employed to predict EV charging demand. The target area is divided into a grid, and EV charging demand is estimated for representative cells. Data is scraped from the Google Maps dataset to predict charging demand.
    % \item We formulate a maximum coverage model to identify optimal charging locations, ensuring all demand values are met. We also determine the capacity of each charging station.
    \item A maximum coverage model is formulated to identify optimal charging locations, ensuring that all demand values are met. The capacity of each charging station is determined.
    % \item We project the 8500-bus grid onto a spatial map and identify the nearest hub for each charging station. Using OpenDSS simulations, we perform power flow calculations to ensure that the proposed capacities in the maximum coverage model are feasible within the power grid network.  
    %\item The 8500-bus grid \cite{8500bus} is projected onto a spatial map, and the nearest hub for each charging station is identified. Power flow calculations are performed using OpenDSS simulations to ensure that the proposed capacities in the maximum coverage model are feasible within the power grid network.
    \item OpenDSS simulations and power flow calculations are performed to validate the capability of the distribution network to accommodate the additional EV charger loads. We conduct power flow analyses both
before and after the integration of new chargers
\end{itemize}}
The rest of the paper is arranged as follows. Section \ref{sec:methodology} discusses the prediction model, maximum coverage model, and algorithm outline. Section \ref{sec:numerics} presents the numerical implementation on a case involving the Dallas Fort Worth (DFW) area and concluding remarks. 
\section{Methodology}\label{sec:methodology}
% In this section, we outline our methodology, beginning with the prediction model to determine charger demand. We then explain the details of our maximum coverage model.% and conclude with OpenDSS simulations.
\blue{This section outlines the methodology for establishing new charging stations. We begin by presenting the prediction model, used to determine charger demand in Phase 1. Phase 2 involves solving the maximum coverage model over an integrated power grid. Finally, we ensure that the newly established chargers do not violate any power grid metrics.}
% \begin{itemize}
%     \item First we select the area under consideration.
%     \item Obtain the location for the charger. 
%     \item Obtain the point of interest near the charger. 
%     \item From the existing traffic situation, obtain the charging demands at the charging stations. 
%     \item Divide the area into a grid. 
%     \item For each grid, obtain the number of points of interest, for example, cafe, schools, grocery, gas stations, theatres etc. 
%     \item Once we have all the POIs, we count them. Then try to relate each number of POI count with the charging demand values. 
%     \item Next, we collect the grids with the higher demands for charging. 
%     \item By solving a facility location problem, we then solve for finalizing the locations of a new candidate charger to establish. 
%     \item Once we have the location decided, we then solve the allowable capacity to be installed there from the DSS simulations. Here we do not want to overload the power grid.
% \end{itemize}
\subsection{\blue{Phase I}: Estimation of EV charging demand}
A primary objective here is to identify the EV charging demand within a specified area. We start by analyzing the relationship between EV charger distribution and nearby amenities, defined as points-of-interest (POIs).%, within the Dallas area (see Fig. \ref{fig:locpoints}). 
 This analysis employs a data-driven approach, leveraging machine learning models to identify key patterns and needs.

\blue{First we finalize the region and then chargers within this region are identified through Google Maps, excluding stations with zero or unknown port counts. In constructing the dataset, the necessary data is extracted, cleaned, and further balanced using sampling techniques. All potential EV charging station candidates are selected for the training process. The EV charger capacity at each potential location is predicted using the extreme gradient boosting (XGB) \cite{chen2016xgboost} model. After the model is fitted to the training data, its performance is assessed using two key metrics: Mean Squared Error (MSE) and $R^2$ score. Further details with the features and the dataset used, please see Section \ref{sec:numerics}.}

\subsection{\blue{Phase II}: Maximum coverage problem formulation}
% The goal here is to determine the optimal routes from demand points to nearby charging stations. 
For any given demand point, % (represented by a red triangle in Fig. \ref{fig:evloc}),
there are two possible choices. The first option is to go to a nearby charging station where an additional charging port can be added at a lower cost per port. The second option is for the demand point itself to become a charging station, thereby meeting both its own demand and that of nearby points. To achieve this, we define four sets of decision variables.

\textit{Decision variables:} For convenience, we define $\mathcal{J}$ as the set of all potential locations of establishing a charger and existing chargers, where $\mathcal{J}:= \mathcal{J}_1 \cup \mathcal{J}_2$. Here, $\mathcal{J}_1$ is the set of all candidate locations for establishing new chargers, and $\mathcal{J}_2$ is the set of all existing chargers. $\mathcal{P}$ is the set of all buses within the 8500 power grid \cite{8500bus}. A binary variable $y_j$ is 1 when a charging station is established at a new site for $j \in \mathcal{J}$. Superscript ``1" ($y^1_j$) denotes a newly opened station, while ``2" stands for an existing one ($y^2_j$). \blue{Decision variable, $x^1_{i,j}$, represents the amount of EV demand traveling from demand point $i$ to charging station $j\in \mathcal{J}$.  Superscripts ``1'' indicate the amount of demand traveling to a newly established charging station $j\in \mathcal{J}_1$, while $x^2_{i,j}$ specifies the demand directed to an existing charging station $j\in \mathcal{J}_2$.} The capacity of newly added ports is represented by two sets of continuous non-negative decision variables: $z^1_{j}$ for a newly opened charging station and $z^2_{j}$ for expanding an existing one.

\textit{Objective function:} Establishing a brand new charging station is more costly than expanding the existing one. Specifically, $term-1$ in equation \eqref{eq:obj}, $E^1$ is greater than $E^2$. 
\blue{For the DFW area, the establishment cost is \$50,000. Values $E^1$ and $E^2$ are adjusted and normalized for the multiobjective objective function in \eqref{eq:obj}. Next, $term-2$ represents the additional capacity required at a charging station, quantified by the number of charging ports. In this context, $C^1$ is the per-port installation cost for Level 2 charging stations. This cost is approximately \$3,000, considering labor, materials, permits, and taxes. Another critical factor is the distance matrix, represented by $C^2_{i,j}$ and $C^3_{i,j}$, obtained by calculating the Euclidean distances between respective nodes.} After that, $term-4$  focuses on the distributed placement of charging stations across the 8500 power grid buses to prevent overloading at any single bus, where the distance matrix $C^4_{p,j}$ represents the distance between bus $p \in \mathcal{P}$ and charging station $j \in \mathcal{J}$. Finally, $term-5$ determines the placement of chargers to ensure that voltage limits for the power grid are maintained, prioritizing buses with higher values of minimum voltage levels. Matrix $C^5_{p,j}$ captures the priority list of preferring bus $p\in\mathcal{P}$ for any charger $j\in \mathcal{J}$. Following is the complete objective function.
\begin{align}\label{eq:obj}
    &\hspace{-0.0cm}\underbrace{\sum_{j\in \mathcal{J}_1}{E^{1}}y^1_j + \sum_{j\in \mathcal{J}_2}{E^{2}}y^2_j}_{term-1} + \underbrace{{C^1}\left(\sum_{j\in \mathcal{J}_1}z^1_{j} + \sum_{j\in \mathcal{J}_2}z^2_{j}\right)}_{term-2} \nonumber \\ &\hspace{-0.0cm}+\underbrace{\sum_{i\in \mathcal{J}_1}\left(\sum_{j\in \mathcal{J}_1}C^2_{i,j}x^1_{i,j} + \sum_{j\in \mathcal{J}_2}C^3_{i,j}x^2_{i,j} \right)}_{term-3} + \underbrace{\sum_{p \in \mathcal{P}}\sum_{j \in \mathcal{J}}C^4_{p,j}y^3_{j}}_{term-4} \nonumber \\ &+ \underbrace{\sum_{p \in \mathcal{P}}\sum_{j \in \mathcal{J}}C^5_{p,j}y^4_{j}}_{term-5}.
\end{align}

\textit{Set of constraints:} First constraint set ensures demand satisfaction, meaning that the charging demand of EVs at any point must be met. %, either by traveling to a newly opened charging station or an existing one.%, where demands $d_i$ for $i \in \mathcal{J}_1$ determined earlier by prediction model (in Stage I).
\begin{align}\label{eqn:constr1}
    \sum_{j\in \mathcal{J}_1}x^1_{i,j} +\sum_{j\in \mathcal{J}_2}x^2_{i,j} \geq d_i \quad \text{ for all } i \in \mathcal{J}_1.
\end{align}
The second set of constraints ensures the capacity limits are not exceeded at either newly opened or existing stations.
\begin{align}\label{eqn:constr2}
    & \sum_{i \in \mathcal{J}_1}x^1_{i,j} \leq z^1_j \ y_j^1\quad \text{ for all } j \in  \mathcal{J}_1 \\
    & \sum_{i \in \mathcal{J}_1}x^2_{i,j} \leq z^2_j \ y_j^2 \quad \text{ for all } j \in  \mathcal{J}_2.
\end{align}
The third set of constraints imposes strict capacity limits (with $Z^1_{\text{cap}}, Z^2_{\text{cap}} >0 $), formulated as follows:
\begin{align}\label{eqn:constr3}
    & z^1_j \leq Z^1_{\text{cap}} \  \ \text{ for all } j \in \mathcal{J}_1, \text{ and }
 z^2_j \leq Z^2_{\text{cap}} \ \ \text{ for all } j \in \mathcal{J}_2. 
\end{align}
There are limits on the maximum number of chargers that can be opened ($ {J}_{max}>0$) within area, in our model:
\begin{align}\label{eqn:constr4}
\sum_{j\in \mathcal{J}}y^1_{j}  \leq {J}_{max}.
\end{align}
The final set of constraints involves the relationships among $y_j^1, y_j^2, y_j^3,$ and $y_j^4$, along with non-negativity and binary constraints on the decision variables. These constraints are omitted here due to space limitations. Our methodology is summarized in flow diagram below. 

\begin{figure}[thb!]
    \centering
    \includegraphics[width=0.81\linewidth]{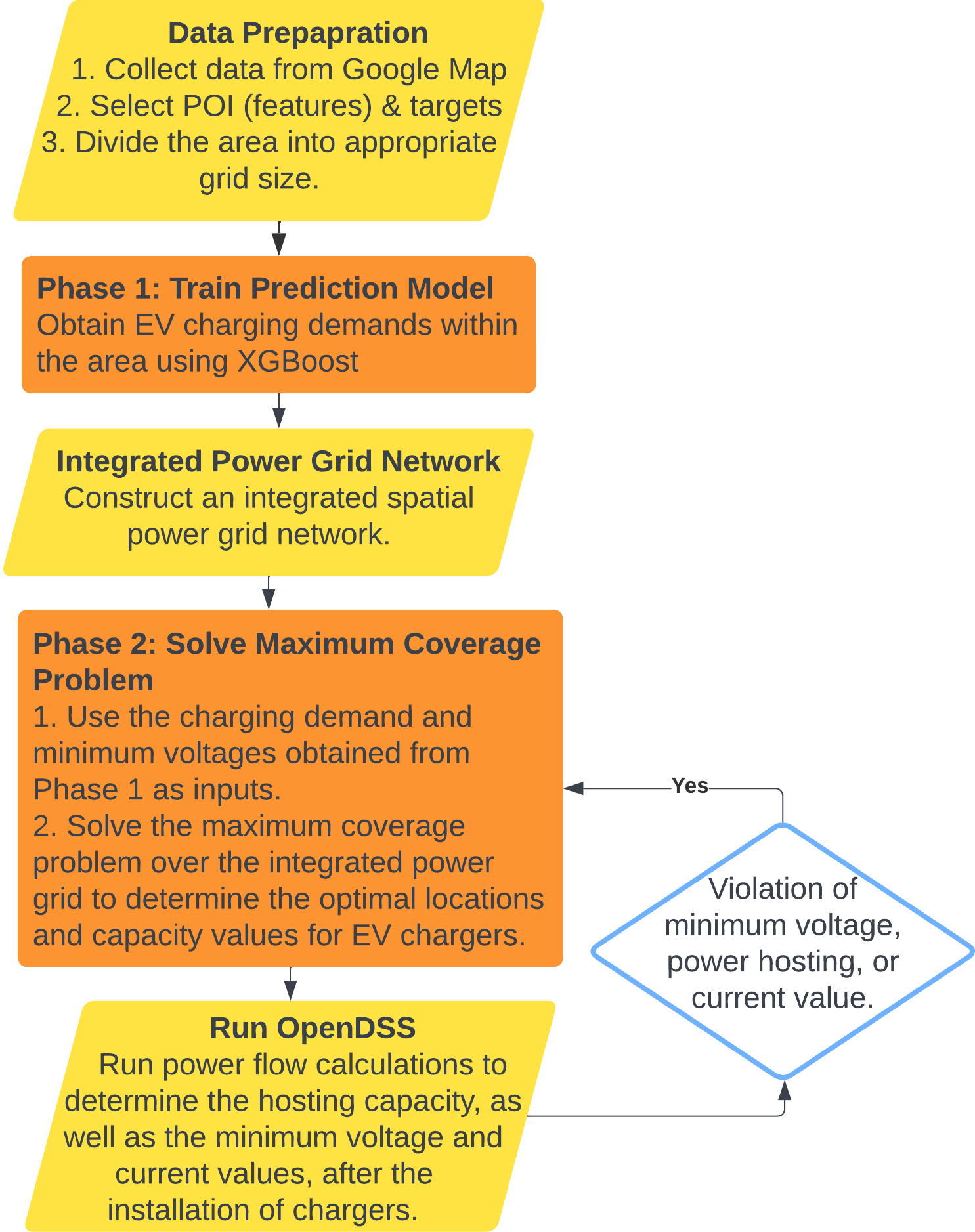}
    % \vspace{-0.45cm}
    \caption{\blue{Flow diagram: EV Charger Infrastructure Planning over an Integrated Power Grid Network.}}
    \label{fig:algorithm}
\end{figure}

\begin{figure}[thb!]
    \centering
    \includegraphics[width=0.9\linewidth]{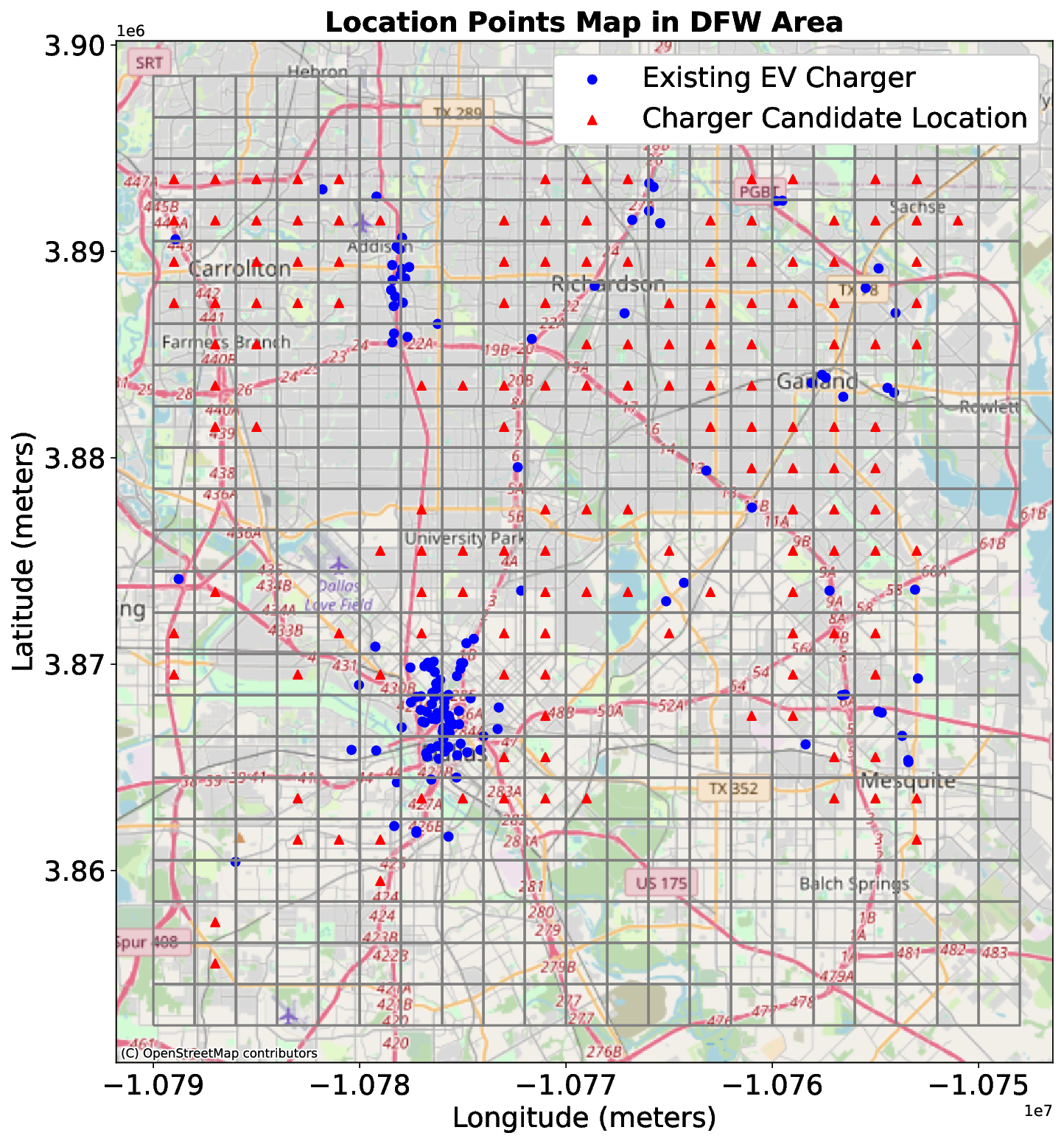}
    \vspace{-0.45cm}
    \caption{Existing and potential EV charger locations in the DFW area}
    \label{fig:evloc}
\end{figure}

\blue{The flow diagram of the proposed approach is depicted in Fig. \ref{fig:algorithm}. Initially, raw data is scraped from Google Maps and processed for Phase 1, which focuses on prediction using XGB. 
After this Phase 1, an integrated spatial-power grid network is constructed and divided into cells. This network includes EV demands at specific cells, existing EV chargers, potential candidate locations for new chargers (Fig. \ref{fig:evloc}), and 8500 buses along with their respective voltage and current values (Fig. \ref{fig:opt_sol}). After solving the maximum coverage problem in Phase 2, the optimal charger locations and sizes are identified. Subsequently, OpenDSS power flow calculations are conducted to ensure that the grid is not overloaded and that the minimum voltage and current values remain within the permissible range.}

% \begin{algorithm}
% \caption{Optimal EV Charger Infrastructure Planning}\label{algorithm}
% \begin{algorithmic}[1]
% % \While{$q_i > 0$ for $i \in \mathcal{D}$}
% \State {\bf Data preparation}
% \Statex \hspace{.36cm}Select POIs (features) \& chargers ports (target).
% \Statex \hspace{.36cm}Divide the area into the appropriate grid size.
% \State {\bf Stage I: Train prediction model XGBoost}
% \Statex \hspace{.36cm}Obtain the EV charging demands within the area.
% \State \textit{Construct an integrated spatial and power grid network.}
% \State \textit{Run OpenDSS power flow calculations for the minimum voltages allowed at buses.}
% \State {\bf Stage II: Solve maximum coverage problem}
% \Statex \hspace{.27cm} Input: EV charging demands from Stage I.
% \Statex \hspace{.27cm} Input: minimum voltages at different buses.
% \Statex \hspace{.27cm} Input: maximum allowed chargers in area.
% \Statex \hspace{.27cm} Output: EV charger infrastructure plan.
% \end{algorithmic}
% \end{algorithm}

\begin{figure} [t]
    \centering
    \includegraphics[width=0.9\linewidth]{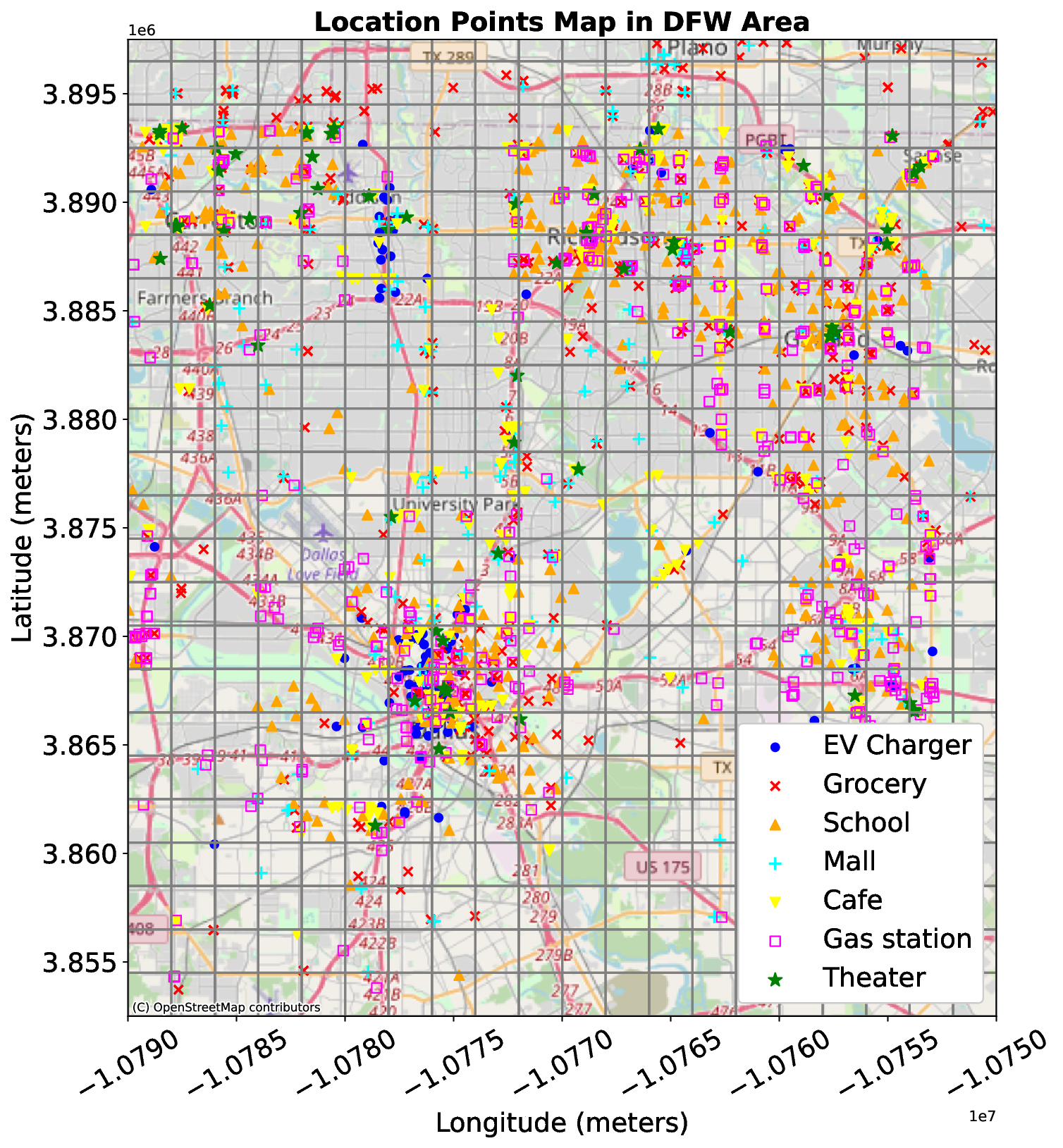}
    \vspace{-0.45cm}
    \caption{Spatial distribution of EV chargers and POIs in the DFW area}
    \label{fig:locpoints}
\end{figure}

\section{Numeirical Experiment}\label{sec:numerics}
\begin{table}[b!]
% \vspace{-0.54cm}
    \centering
        \caption{Correlation of POI categories with EV charging stations}
    \label{tab:evcs_corr}
    \begin{tabular}{lc}
    \toprule
        POI Category         & Correlation \\ 
    \midrule
        Gas Station          & 0.66        \\ 
        Grocery Store       & 0.72        \\ 
        Cafe \& Restaurant   & 0.69        \\ 
        Shopping Mall       & 0.55        \\ 
        Theater              & 0.56        \\ 
        School               & 0.59        \\ 
    \bottomrule
       \end{tabular}
\end{table}
In this section, the proposed methodology is tested on a real spatial dataset. The DFW area is selected along with the 8500-bus network (see Fig. \ref{fig:locpoints}).
%The case study region is defined by the following geographical boundaries: north at 33.0165, south at 32.6769, east at -96.5688, and west at -96.9282. 
Chargers in this region were identified through Google Maps, excluding stations with zero or unknown port counts. This yields approximately {110} charging stations, predominantly managed by established EV charging companies such as EVgo, Tesla, and Blink, which represent most of the active public chargers in the area. In gathering charger information, we obtained the plus codes, which can be converted to latitude and longitude, as well as the actual number of ports at each station, serving as the target variable for estimating maximum charging capacity. \blue{For additional charger power assessment, a 6kW rating was chosen, aligning with widely used Level 2 AC chargers (3.3kW–7.2kW). This ensures compatibility with residential and commercial infrastructure while maintaining reasonable grid impact. The 6kW assumption provides a balanced estimate of aggregated demand at public charging sites, preventing excessive peak loads while supporting efficient EV charging. }% Many EV models also support this power level, making it a practical benchmark for infrastructure planning.}  
 \blue{For POIs, we incorporated the geographical coordinates of each point in our analysis. Six major categories were selected as potential drivers (features): grocery stores, schools, malls, cafes, gas stations, and theaters, as these are locations where people typically spend extended time, making them viable candidates for EV station placement (see Table I). Residential areas were excluded from our POI analysis, as planning the charging infrastructure for residential areas requires a separate approach (see Fig. \ref{fig:locpoints}). Each 2$\times$2 km$^2$ cell, optimized through a parameter sweep for the best accuracy, serves as a distinct data point in the analysis.}

% \begin{figure}
%     \centering
%     \includegraphics[width=0.9\linewidth]{Figures/correlation_matrix.pdf}
%     \caption{correlation matrix}
%     \label{fig:corr_m}
% \end{figure}

% The POIs and charging stations are illustrated in Fig. \ref{fig:locpoints}, with a superimposed gray grid representing subdivisions. Each cell in the grid, covering an area of 2 km by 2 km, serves as a distinct data point in our analysis. The grid size was optimized through a parameter sweep to achieve the best training accuracy.%, as will be detailed in the subsequent section. 

% For the training and prediction sets, we defined four classifications, summarized in Table \ref{tab:table1}. Classification $C1$ and $C4$ are used for training. $C1$ represents  the relationship between existing charger capacity and its surrounding area within each 2 km cell, while $C4$ represents areas lacking both EV chargers and POIs, which we aim to model through machine learning.  Classification $C3$ includes cells with POIs but no existing EV chargers, indicating areas with potential charging demand. Each $C3$ cell is treated as a potential EV charger location, with the cell center serving as the graphical point for potential placement. The objective is to predict the required charging capacity for these potential locations. Classifications $C2$ are excluded from the study, as they contain EV chargers without nearby POIs, which are outside the scope of interest.
\begin{table}[t]
    % \vspace{-0.54cm}
    \centering
      \caption{Classifications based on EV charger and POI presence}
    \label{tab:table1}
    \begin{tabular}{ccc}%{|c|c|c|}
           \toprule
        Classification & EV Charger & POIs\\
        \midrule
         $C1$ & \ding{51} & \ding{51}\\ 
         $C2$ & \ding{51} & \ding{55}\\ 
         $C3$ & \ding{55} & \ding{51}\\ 
         $C4$ & \ding{55} & \ding{55}\\ 
         \bottomrule
    \end{tabular}\vspace{-.54cm}
\end{table}
\blue{Four classifications were defined for the training and prediction sets, as summarized in Table \ref{tab:table1}. $C1$ (existing charger capacity and surrounding area) and $C4$ (areas without chargers or POIs) were used for training. $C3$ includes cells with POIs but no chargers, treated as potential locations for predicting required charging capacity, while $C2$ (chargers without POIs) was excluded. The training dataset comprises 131 data points, while the prediction dataset includes 352 data points.  Within the training dataset, in order to remove imbalance between $C1$ and $C4$ and in order to address this imbalance between $C1$ and $C4$, the Synthetic Minority Over-sampling Technique (SMOTE) was applied. The training target is the charging capacity of existing stations, represented by the total port number, and the training features are the {number} of existing POIs, represented as binary indicators.}

\blue{To train and evaluate the model, the resampled training dataset is split into training and testing subsets, and the XGB \cite{chen2016xgboost} is used for prediction. A comparative analysis with other methods showed that {XGB} provides higher accuracy. Since no significant spatial relationships among the features were found, the data is treated as non-sequential. As extensively discussed in the literature, the spatial dependencies between features are difficult to establish when predicting EV charger demands. Given the structured, non-sequential nature of the data, and leveraging the boosting capabilities of XGB, it was selected for this phase. Key hyperparameters were set, including 500 boosting rounds, a maximum tree depth of 8, a learning rate of 0.01, a training instance sampling rate of 0.7, and a feature sampling rate 0.8. Table \ref{tab:performance_metrics} shows the model’s performance.}

%The training process begins by selecting all potential EV charging station candidates within classification $C3$, represented as red triangles in Fig. \ref{fig:evloc}. Using the XGB\cite{chen2016xgboost} model, the EV charger capacity at each potential location is predicted. %The training results demonstrate promising accuracy, as shown in Fig. \ref{fig:train}. 
%After fitting the model to the training data, predictions were generated for both the training and test sets. Model performance was assessed using two key metrics: Mean Squared Error (MSE) and $R^2$ score. 
% Table \ref{tab:performance_metrics} shows the model’s performance with Training MSE, Training $R^2$, Test MSE, and Test $R^2$ values for easy comparison. This setup allows us to evaluate the XGB model’s ability to generalize and predict charging capacity effectively.

% \begin{figure}[thb!]
%     \centering
%     \includegraphics[width=0.9\linewidth]{Figures/trainingresults.pdf}
%     \vspace{-0.45cm}
%     \caption{Actual and Predicted Charging Capacity with Residual Analysis}
%     \label{fig:train}
% \end{figure}

\begin{table}[thb!]
    \vspace{-0.09cm}
    \centering
      \caption{Performance metrics for the model}
    \label{tab:performance_metrics}
    \begin{tabular}{lcccc}
        \hline
        Method & Training MSE & Training $R^2$ & Test MSE & Test $R^2$ \\
        \hline
        XGB    &   0.8919   &  0.9826 & 1.1876&  0.9792\\
        % XGB & 0.7690 & 0.9846 & 0.6510 & 0.9817 \\
        \hline\\
    \end{tabular}\vspace{-.63cm}
\end{table}
% \subsubsection*{Integrated spatial and power network}
% \begin{figure}[t]
%     \centering
%     \includegraphics[width=0.9\linewidth]{Figures/Charg_stn_8500.pdf}
%     \vspace{-0.45cm}
%     \caption{Location Points Map}
%     \label{fig:ev_cand_8500}
% \end{figure}

\begin{figure}[thb!]
    \centering
    \includegraphics[width=0.9\linewidth]{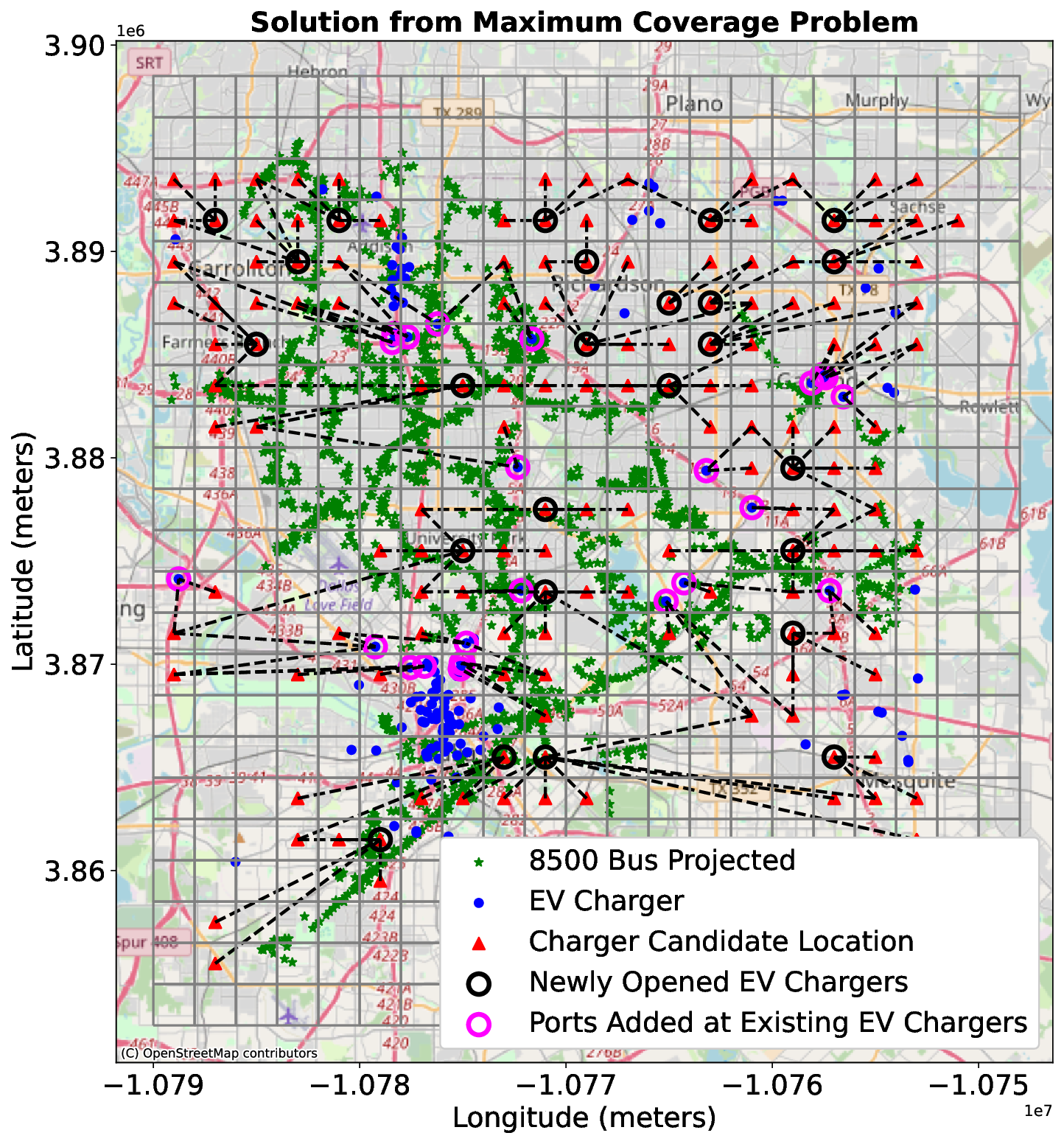}
    \vspace{-0.45cm}
    \caption{Optimal EV charger placement, considering the 8500-bus power grid.}
    \label{fig:opt_sol}
    % \vspace{-0.27cm}
\end{figure}

From the power grid perspective, in this study, we use the IEEE 8500-node distribution test feeder \cite{8500bus} as a case study to build an integrated network for our analysis. Our analysis focuses on the primary level of the distribution network. We convert the primary distribution network's local coordinates into real geodesic coordinates and use the DFW region (shown in Fig. \ref{fig:opt_sol}). The primary level of the 8500-node feeder is then mapped onto this spatial framework. Finally, we identify the nearest buses within the 8500-bus system to the spatial nodes by minimizing Euclidean distances. Only buses with a sufficient margin above the lower voltage limit are considered in the candidate pool for optimal planning.

\begin{figure} [t]
    % \textcolor{blue}{
    \centering
    \includegraphics[width=0.9\linewidth]{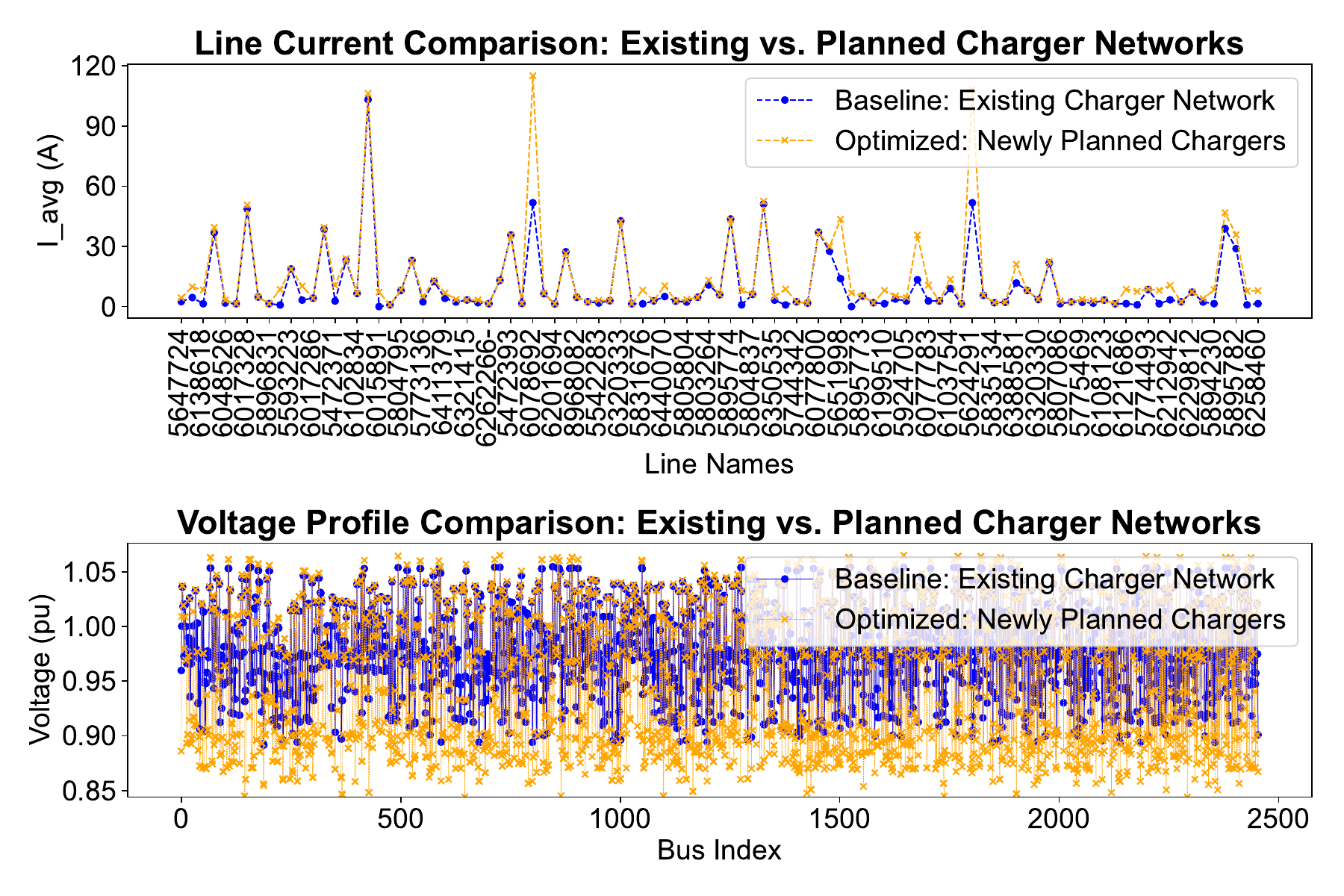}
    \vspace{-0.45cm}
    \caption{\blue{Impact of Charger Network Optimization on Line Current and Voltage Profiles in an 8500-Node Distribution Grid.}}
    \label{fig:grideval}
    % }
\end{figure}

% We map the locations of existing chargers, along with their rated power, onto the 8500-node distribution test feeder. 
A snapshot power flow analysis of the network at its peak load capacity is performed using OpenDSS. \blue{To validate the capability of the distribution network to accommodate the additional EV charger loads, we conducted power flow analyses both before and after the integration of new chargers (see Fig. \ref{fig:grideval}). 
The charging stations are modeled as energy storage units with charging profiles and rated capacities based on the data from the DFW area. For peak load power flow analysis, where time-series variations are not considered, the EV charging stations are modeled as static loads rated at their maximum capacity providing a conservative assessment of the grid's ability to handle peak EV charging demand under stressed conditions.
Voltage levels at buses are evaluated to ensure they remain within permissible limits, with the lower limit set at 0.8 per unit, common in baseline distribution network feeders. Additionally, we monitored the current through distribution lines, which serves as a measure for assessing thermal limits. However, because OpenDSS assigns a default current rating of 600 A to all lines, regardless of network configuration or line type, it is not possible to confirm whether the current flow remains within the actual maximum permissible limits.} 
% Only buses with a sufficient margin above the lower voltage limit are considered in the candidate pool for optimal planning. These selected buses are then mapped onto the spatial network and incorporated into the optimal planning framework. Fig. \ref{fig:opt_sol} presents the optimal solution obtained from the case study.% Newly established chargers are marked with black circles, while additional ports added to existing chargers are indicated by magenta circles. %In this solution, most charger placements were determined with consideration for power grid constraints.
% \vspace{9cm}
% \blue{Next,  we conduct power flow simulations using OpenDSS onto the solutions obtained from Phase 2 (Fig .\ref{fig:opt_sol}) to ensure they do not overload the 8500-grid \cite{8500bus}. Results, shown in Fig. \ref{fig:grideval}, compare the baseline (existing chargers) with the optimized scenario (new chargers) based on voltage and current profiles. The top plot shows the average current (I\_avg) across all lines, confirming that thermal limits are respected in the optimized scenario. The bottom plot compares voltage profiles, ensuring all bus voltages remain above the critical 0.80 pu threshold, demonstrating voltage stability. }
\blue{These results validate that the optimized charger placement meets both thermal and voltage constraints while balancing increased demand and operational reliability.}

\section*{Conclusion}
% \label{sec:conclusion}
This paper introduces a \blue{two-phase approach for the strategic deployment of EV chargers by integrating spatial statistics and maximum coverage analysis over an integrated spatial power grid. By respecting capacity constraints from the distribution grid perspective, this approach prevents overloading and inefficiencies in the power network.} It offers a sustainable pathway for cities to expand EV infrastructure, supporting the growth of a reliable and well-balanced charging network that aligns with the power grid's capacity. Future research could focus on adapting this strategy to various urban and suburban settings, further enhancing the integration of transportation and power networks for widespread EV adoption.

\bibliography{Gen_meet.bib}
% \vspace{12pt}
\end{document}